\documentclass[conference]{IEEEtran}
\IEEEoverridecommandlockouts
\usepackage{cite}
\usepackage{amsmath,amssymb,amsfonts}
\usepackage{graphicx}
\usepackage{textcomp}
\usepackage{xcolor}
\usepackage{hyperref}
\usepackage{float}
\usepackage{booktabs}
\usepackage{array}
\usepackage{soul}
\usepackage[compatibility=false]{subcaption}
\usepackage{enumitem}
\usepackage{algorithm}
\usepackage{algpseudocode}
\usepackage{todonotes}

\usepackage{balance}

\def\BibTeX{{\rm B\kern-.05em{\sc i\kern-.025em b}\kern-.08em
    T\kern-.1667em\lower.7ex\hbox{E}\kern-.125emX}}

\usepackage[dvipsnames]{xcolor} 

\usepackage{adjustbox}
\usepackage{mdframed}   
\usepackage{tabularx}

\usepackage[table]{xcolor}

\usepackage{tikz}

\renewcommand{\figureautorefname}{Fig.}
\renewcommand{\thefigure}{\arabic{figure}}

\definecolor{codehighlight}{HTML}{eaf3ff} 

\definecolor{coderemove}{HTML}{ffecec} 
\definecolor{codeadd}{HTML}{eaffea}    

\newcommand{\pietro}[1]{{\textcolor{cyan} {Pietro: \textbf{#1}}}}

\usepackage[most]{tcolorbox}
\usepackage{setspace}

\definecolor{titlegray}{RGB}{80,80,80}
\definecolor{bodygray}{RGB}{245,245,245}

\newenvironment{mybox}[1]{%
  \begin{tcolorbox}[
    enhanced,
    title=#1,
    colback=bodygray,
    colframe=black,
    colbacktitle=titlegray,
    coltitle=white,
    fonttitle=\bfseries,
    boxrule=1pt,
    arc=4pt,
    left=6pt, right=6pt, top=6pt, bottom=6pt,
  ]
  \setstretch{0.95}
}{%
  \end{tcolorbox}
}

\usepackage{tikz}
\usepackage{scalerel}

\usepackage[subtle,paragraphs=normal,tracking=normal]{savetrees}

\begin{document}

\raggedbottom


\title{Will It Break in Production? Metric-Driven Prediction of Residual Defects in Python Systems}

\author{
\IEEEauthorblockN{Giuseppe De Rosa\IEEEauthorrefmark{1}\IEEEauthorrefmark{2}, Pietro Liguori\IEEEauthorrefmark{1}}
\IEEEauthorblockA{\IEEEauthorrefmark{1}\textit{University of Naples Federico II}, Naples, Italy \\
giuseppe.derosa20@unina.it, pietro.liguori@unina.it}
\IEEEauthorblockA{\IEEEauthorrefmark{2}\textit{IMT School for Advanced Studies Lucca}, Lucca, Italy \\
giuseppe.derosa@imtlucca.it}
}

\maketitle
\thispagestyle{plain}
\pagestyle{plain}

\begin{abstract}
Python's dynamic nature complicates testing and increases the possibility that some defects evade detection, so an effective fault prediction becomes essential. We examine whether post-release faults can be predicted using modern ML and DL. Using a balanced dataset of over 4,000 labeled faults with 83 product, process, statistical, and Python-specific metrics plus normalized code representations, we conduct cross-project experiments. LLMs and unsupervised models fail to distinguish residual from non-residual faults, while supervised metric-based models (RandomForest, XGBoost, CatBoost) perform far better,  yielding a 0.85-0.9 recall and cutting false negatives by an order of magnitude. Process metrics, especially age, churn, and developer-activity, alongside class and file size, consistently prove most predictive. Notably, the Principal Component Analysis shows that metrics and code embeddings occupy distinct regions of the representation space, suggesting that they capture complementary rather than redundant information.
\end{abstract}


\section{Introduction}
\label{sec:intro}
Software failures continue to impose severe operational and economic consequences across modern digital infrastructures. 
From the \$460 million trading loss at Knight Capital~\cite{dolfing2019knightcapital} to large-scale transportation disruptions and safety-critical incidents~\cite{cnn2018uber, ieee2019boeing737}, the inability to prevent defects from reaching deployment environments remains a central challenge for software dependability. 
Industry analyses echo this urgency: according to Gartner, organizations that invest in ``digital immunity", i.e., the systematic detection, containment, and mitigation of defects, can reduce downtime by up to 80\%, significantly improving resilience and service continuity~\cite{gartner_digital_immunity_2023}.  
Accurately identifying fault-prone components before release, therefore, plays a decisive role in reducing maintenance effort, improving testing strategies, and increasing trust in the ever-growing ecosystem of software-driven systems.

Despite substantial advances in defect prediction, one category of faults continues to elude even rigorous testing pipelines: \emph{residual faults}, i.e., defects that remain unnoticed until the software is already in production.  
In the reliability literature, these are also referred to as \emph{post-release faults}, since the defining characteristic of a residual defect is precisely that it survives all pre-release verification activities and manifests only after deployment~\cite{nagappan2006mining, rwemalika2019industrialstudy}.  
Using post-release failures as an operational definition is, therefore, standard practice: a defect revealed in the field is, by construction, one that the testing and quality-assurance processes failed to expose.

These faults are uniquely problematic. 
They often surface only under realistic operational conditions that are difficult or impossible to reproduce in controlled test suites; they tend to arise from subtle interactions, rare execution paths, or latent corner cases; and, critically, they rarely exhibit clear syntactic or structural patterns that distinguish them from non-residual defects.  
This makes predicting which faults will escape pre-release detection intrinsically more challenging than predicting fault-proneness in general. 

Recent research in machine learning (ML) and deep learning (DL) for defect prediction has shown promising results by leveraging combinations of product and process metrics, along with learned code representations~\cite{giray2023use}.  
Yet several long-standing issues remain unresolved, and they become even more critical when the target is residual faults rather than generic defect-proneness.  
First, cross-project generalization is notoriously weak: models trained on one project frequently underperform when applied to others~\cite{chowdhury2024situ}. This limitation is exacerbated by the rarity of residual faults and their project-specific characteristics.  
Second, DL approaches often lack interpretability, limiting their usefulness in safety-critical settings where understanding why a component is predicted as risky is as important as the prediction itself~\cite{giray2023use, cinque2025cosmos}.  
Third, residual faults are inherently rare, creating a challenging class imbalance that distorts learning and inflates apparent performance~\cite{hall2011systematic, catal2009systematic}.  
These limitations motivate the need for carefully designed datasets and a broad spectrum of software metrics to capture language-specific behaviors that may influence residual fault manifestation.

This paper investigates residual fault prediction. Predicting residual faults is not equivalent to standard defect prediction: traditional defect predictors estimate whether a defect exists in a component, implicitly conflating two statistically different populations: faults that will be caught during testing and faults that will escape into production. We formulate a distinct prediction target: estimating the probability that a defect, once introduced, survives all pre-release validation activities and reaches production. This distinction is operationally significant because it enables targeted interventions such as selective manual review or fuzzing only for components with high escape risk rather than the entire codebase.
To support this task, we leverage a diverse range of traditional software metrics, including product metrics that quantify size, control-flow complexity, documentation quality, coupling, and inheritance structures; Python-specific syntactic indicators; statistical dispersion measures; and process metrics that capture code churn, temporal evolution, commit activity, and developer behavior. We focus on Python because of its widespread adoption in modern software infrastructures and its dynamic, interpreted nature, which makes defects more likely to manifest only at runtime~\cite{tiobe_index}. These characteristics increase the probability that subtle faults escape pre-release testing, making Python a particularly relevant and challenging setting for studying residual defects~\cite{rezaalipour2024empiricalstudy, widyasari2022real}.


Our study yields three key contributions:

\begin{enumerate}

\item \textbf{A new dataset of residual and non-residual Python faults.}  
We build a balanced and reproducible dataset of more than 4,000 classified faults from real-world Python projects, complemented by 500 additional cases from BugsInPy. Each instance includes code, software metrics, and version-control information, enabling systematic experimentation on residual-fault prediction and supporting a repeatable evaluation.

\item \textbf{A systematic assessment of LLM-based code representations for residual fault prediction.}  
We evaluate both proprietary and fine-tuned open LLMs under uniform conditions. Although their embeddings contain meaningful signals, the models themselves perform unreliably: closed-source systems achieve F1-scores between 0.43 and 0.72, often collapsing into degenerate behavior, while fine-tuned models reach only 0.35–0.38 and yield a recall of 0.3 on average. A representation-space analysis shows that embeddings remain largely orthogonal to metrics (maximum correlation 0.27; mean 0.24), indicating that LLMs encode complementary but insufficient information for this task.

\item \textbf{An investigation of software metrics through lightweight ML models.}  
We assess how far one can go using only traditional software metrics and simple numerical models. Unsupervised anomaly detectors perform poorly, confirming that residual faults are not statistical outliers. In contrast, supervised learners such as RandomForest, XGBoost, and CatBoost achieve F1-scores of 0.71–0.73 and reach 0.85-0.9 recall, reducing false negatives by an order of magnitude compared to LLMs. Feature-importance and SHAP (i.e., a method that assigns each feature a quantitative contribution to the model’s prediction) analyses highlight process, size, and developer-activity metrics as the most informative families, revealing stable structural and historical patterns that support effective residual-fault prediction.
\end{enumerate}

To foster both academic and industrial research on this topic, we publicly release our replication package~\cite{RP}, which includes the datasets and scripts used for the analyses in the paper.

The remainder of this paper is organized as follows. 
Section~\ref{sec:prob_statement} introduces the background and motivates our focus on residual faults. 
Section~\ref{sec:methodology} details the data collection, metric extraction, and model design. 
Section~\ref{sec:setup} describes the experimental setup, while Section~\ref{sec:evaluation} presents our evaluation results. Section~\ref{sec:discussion} discusses the results.
Related work is discussed in Section~\ref{sec:related}, and threats to validity are outlined in Section~\ref{sec:threats}. 
Finally, Section~\ref{sec:conclusion} concludes with reflections and directions for future work.

\section{Background}
\label{sec:prob_statement}

\textit{Software defect prediction} aims to proactively identify modules within a system that are likely to contain undiscovered bugs before these defects manifest during real-world operation. Typically, this involves collecting diverse code metrics, such as complexity, size, object-oriented design indicators, change history, and developer-related factors, and applying machine learning algorithms to detect patterns associated with defect-prone components~\cite{rathore2019study}. Conventional approaches train classification models on historical defect data, allowing development teams to prioritize testing and debugging efforts~\cite{thota2020survey}.

However, most existing prediction techniques focus primarily on defects identified during pre-release testing, neglecting those that surface only after deployment. These post-release issues, commonly referred to as \textit{residual defects}, exhibit fundamentally different characteristics. They often affect components that appeared stable during testing, but fail in production under atypical inputs, varied deployment contexts, or unanticipated runtime interactions. Because of their contextual and environment-dependent nature, residual defects tend to be harder to detect, more costly to resolve, and more disruptive when uncovered in operational settings~\cite{rwemalika2019industrialstudy}.

Residual defects are especially concerning in systems developed using dynamically typed languages. Among them, Python has experienced explosive growth across a wide range of domains, becoming foundational in web development (e.g., Django, Flask), machine learning (e.g., TensorFlow, scikit-learn), and data analytics (e.g., Pandas, NumPy~\cite{github2024octoverse, realpython2024news}. While its flexibility and expressive syntax contribute to its popularity, Python’s dynamic typing and interpreted execution model make it particularly susceptible to runtime errors, such as type mismatches, undefined variable access, and improper exception handling, that are difficult to detect via static analysis or conventional testing~\cite{nanz2015study}. This susceptibility introduces significant operational risk, especially in large-scale and distributed environments~\cite{cisq2022}. 




\section{Methodology}
\label{sec:methodology}
\begin{figure*}[ht] 
    \centering
    \includegraphics[width=2\columnwidth]{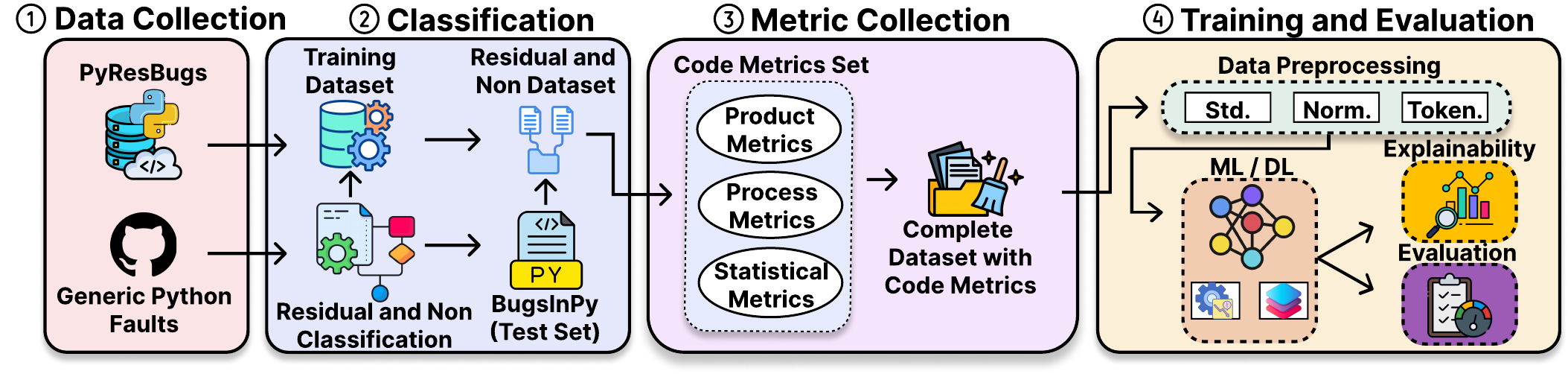} 
    \caption{Detailed methodology adopted in this work.}
    \label{fig:methodology}
    \vspace{-3mm}
\end{figure*}

This section describes the methodology used to predict residual faults. \figurename{}~\ref{fig:methodology} presents an overview of the proposed workflow, which consists of four main stages:
(i) data collection from PyResBugs and real-world Python projects;
(ii) classification of each fault as residual or non-residual to build the training and test sets;
(iii) collection of product, process, and statistical software metrics for every instance; and
(iv) training and evaluation of both ML and DL models, including preprocessing and explainability analyses.

\subsection{Data Collection}
\label{subsec:data_collection}
The success of fault prediction models largely depends on the quality, representativeness, and balance of the underlying datasets. Our data collection process carefully addresses these concerns, systematically gathering examples of both residual (post-release) and non-residual (pre-release) faults.

For residual faults, we leveraged the publicly available \textit{PyresBugs} dataset, which contains $5,007$ documented post-release bugs extracted from 76 prominent Python frameworks and libraries~\cite{cotroneo2025pyresbugsdatasetresidualpython}. Residual faults are bugs that escape pre-release testing and only surface after the software is deployed. PyresBugs was initially constructed by aggregating faults from two primary sources. First, faults were curated from eleven existing datasets published in major software engineering conferences and journals over the past five years. This step ensured broad coverage of well-established, community-validated fault examples. Second, the dataset was augmented with additional categories of residual faults, specifically targeting those challenging to identify in standard testing phases, such as time-related issues and concurrency errors.
To reliably identify genuine residual faults, we employed a multi-stage filtering procedure: bug reports from software repositories (e.g., GitHub) were analyzed to identify defects explicitly discovered post-release. Each identified defect was confirmed by examining associated commits to verify the availability of code-level fixes, enabling the extraction of both faulty and corrected code versions. Furthermore, we verified that each fault originated from source-code defects (within method-level), thus excluding cases involving documentation issues, configuration errors, or build system problems.

To construct the set of non-residual faults, we mined a large number of Python projects selected to resemble those included in PyResBugs in terms of size, maturity, and development activity. This alignment is essential to avoid systematic differences between the two defect categories that might arise from project-specific conventions. For each repository, we traversed the entire commit history and applied a keyword-based heuristic to identify commits related to defect correction. Following established practice in mining software repositories, we flagged a commit as bug-related if its message contained terms such as fix, fault, bug, defect, crash, issue, and so on. Keyword-based identification of bug-fixing commits is a standard approach in large-scale repository mining, and although imperfect, it offers a good balance between coverage and precision when the goal is to gather a broad sample of natural faults~\cite{zafar2019towards, stanislav2017boosting}. All bug-fixing commits discovered in this way were retained, and each associated defect instance was marked as a possible buggy commit.

The purpose of this collection was to gather a broad and diverse set of naturally occurring bugs. For each qualifying commit, we recorded project metadata, commit ID, file paths, the corresponding diff patches, the corresponding faulty version, and the issue or pull request ID. These patches provide sufficient information to characterize the fault and compute all product, process, and statistical metrics used in this study. The resulting collection forms a large corpus of authentic, possibly buggy commits. The next step will be to clean and classify them as non-residual faults by inspecting the associated issue.

\subsection{Residual or Non-Residual Classification Task}

\begin{algorithm}[ht]
\caption{Classify a commit as residual or non-residual}
\label{alg:classify-commit} 
\begin{algorithmic}[1]
\Procedure{ClassifyCommit}{$r$}
  \If{linked issue exists}
    \State collect strong pre/post hints from issue time 
    \If{some strong pre hint} \Return pre-release \EndIf
    \If{some strong post hint} \Return post-release \EndIf
    \State compute soft pre/post scores
    \If{$pre\_score > post\_score$} \Return pre-release
      \ElsIf{$post\_score > pre\_score$} \Return post-release \EndIf
    \If{author is external non-contributor} \Return post-release
      \ElsIf{author is internal} \Return pre-release \EndIf
  \EndIf
  \State $(cd, frd) \gets$ commit date and first stable release date of repo
  \If{$cd \neq \bot$ and $(frd = \bot$ or $cd < frd)$} \Return pre-release \EndIf
  \State \Return unknown
\EndProcedure
\end{algorithmic}
\end{algorithm}

To effectively perform this classification step, we rely on a dedicated heuristic that infers whether each bug-fixing commit resolves a pre-release or post-release fault, using only locally available historical repository information.
The procedure \textsc{ClassifyCommit} detailed in Algorithm~\ref{alg:classify-commit} operates in stages, progressively incorporating increasingly weaker forms of evidence. The algorithm first checks whether the commit is linked to an issue (line~2). If so, it extracts decisive temporal signals from the issue creation time (line~3). Such signals include (i) whether the issue was created before the project's first stable release, which unambiguously identifies it as a pre-release defect, and (ii) whether the issue contains explicit references to behaviour observed in a shipped version, like numeric version identifiers (e.g., \texttt{v1.X}, \texttt{release~X}, \texttt{build~X}), "affects–X.Y'' labels, or regression tags, which indicate that the problem occurred in code already distributed to users. Issues created before the first stable release are immediately classified as pre-release (line~4), whereas those containing evidence tied to a released version are immediately classified as post-release (line~6), without considering weaker heuristics.

When no such hard evidence is available, the algorithm computes two soft scores, \emph{pre\_score} and \emph{post\_score}, each aggregating multiple weak heuristics (line~8). These include milestone information (closed milestones contributing to the post-release score, pre-release qualifiers contributing to the pre-release score), textual indicators in the issue body (internal-test markers increasing the pre-release score, bug-report templates and reproduction steps increasing the post-release score), and the reporter's role (external reporters incrementing the post-release score, contributors incrementing the pre-release score). All soft indicators are boolean and combined through an unweighted sum, with no tuned parameters or learned weights, to ensure reproducibility across projects and to avoid project-specific bias.
The full list of indicators is provided in our replication package~\cite{RP}.

After accumulating the scores, the algorithm compares the two totals.  
Suppose \emph{pre\_score} exceeds \emph{post\_score}. In that case, the majority of temporal and contextual evidence associated with the issue points to activities that typically occur \emph{before} a release, e.g., developers discovering a defect during testing, code review, or stabilization. In other words, a higher \emph{pre\_score} suggests that the bug was fixed before deployment, and the commit is therefore classified as a pre-release fix.  
Conversely, a higher \emph{post\_score} indicates that the evidence aligns more with characteristics of field failures (e.g., reports submitted after deployment, user-facing symptoms, production traces), supporting a post-release classification.

Each score is, by design, an imperfect proxy: for instance, an external reporter may occasionally file an issue discovered through source inspection rather than through use of a deployed version, and textual markers such as reproduction steps can appear in pre-release bug reports as well. The algorithm mitigates this by never relying on a single indicator in isolation; instead, it aggregates all available signals and decides only when the balance of evidence favours one category. When the evidence is insufficient and/or conflicting, the commit is conservatively labeled \emph{unknown} and excluded from subsequent analyses. 
When both scores remain tied, the algorithm falls back on the identity of the issue author (line~12) as a tie-breaker. We noticed that issues reported by external users are far more likely to reflect failures observed in the field: they interact only with deployed versions of the software; therefore, issues reported by non-contributors almost always correspond to failures observed in production. Instead, issues filed by contributors typically originate from internal testing or development before a release. The reporter identity, thus, represents a weak but meaningful signal for distinguishing between defects discovered during pre-release activities and those surfaced only after deployment. This signal is particularly valuable in cases where the temporal evidence extracted from timestamps is insufficient to disambiguate the classification (i.e., when \emph{pre\_score} and \emph{post\_score} are tied). We use the identity of the \emph{issue author} to determine internal vs. external origin: reporters with no prior contributions are treated as external, while reporters with commits or reviews are considered internal.

When no linked issue exists, the algorithm falls back to a timestamp-only rule (line~17): commits dated before the first stable release are labeled pre-release, while all other commits receive the \emph{unknown} label. This could appear as an asymmetry: pre-release can be inferred from timestamps alone, but post-release cannot, because a commit made after a first release may still address a pre-release fault whose fix was delayed. However, it is well-known that faults escaping the testing phase and manifesting in production are far less than the ones surfacing in the testing phase, especially in mature projects with good test suites. As a consequence, the heuristic adopts a more conservative approach with the post-release faults to avoid introducing false positives, which we consider the more damaging threat to validity to the dataset quality.

To assess the accuracy of the heuristic, two authors independently inspected a random sample of 400 classified commits. Each commit was judged against its linked issue, commit message, and surrounding development context. The two raters achieved a Cohen's $\kappa > 0.80$, indicating strong inter-rater agreement, and the heuristic's labels matched the consensus ground truth in approximately 95\% of cases.

\begin{table*}[ht] \caption{Software Metrics Adopted} \label{tab:metrics} \centering \small \begin{tabular}{p{2.5cm}|p{14.5cm}} \toprule \textbf{Category} & \textbf{Metric [Acronym]} \\ \midrule \multicolumn{2}{l}{\textbf{Product Metrics}} \\ \midrule Complexity & \textbf{Method:} Cyclomatic Complexity [\textbf{CC}], Maximum Nesting Depth [\textbf{MND}], Number of Paths [\textbf{NP}], \textit{Halstead:} Difficulty [\textbf{HD}], Length [\textbf{HL}], Vocabulary [\textbf{HV}], Volume [\textbf{HVOL}], Effort [\textbf{HEFF}], Maintainability Index [\textbf{HMI}], Distinct Operators [\textbf{HDOP}], Distinct Operands [\textbf{HDND}], Total Operators [\textbf{HTOP}], Total Operands [\textbf{HTOA}] \textbf{File:} Cyclomatic Complexity [\textbf{F-CC}], Maximum Nesting Depth [\textbf{F-MND}], Number of Paths (log scale) [\textbf{F-NPLOG}] \\ \midrule Size & \textbf{Method:} Lines of Code [\textbf{LOC}], Blank Lines [\textbf{BLOC}], Declaration Lines [\textbf{DLOC}], Executable Lines [\textbf{ELOC}], Statement Count [\textbf{STMT}], Declaration Statements [\textbf{DSTMT}], Executable Statements [\textbf{ESTMT}], Number of Inputs [\textbf{NIN}], Number of Outputs [\textbf{NOUT}], Number of Exits [\textbf{NE}], Number of Early Exits [\textbf{NEE}] \textbf{Class:} Class Lines [\textbf{CLLOC}], Code Lines [\textbf{CCODE}], Declaration Lines [\textbf{CDLOC}], Executable Lines [\textbf{CELOC}], Number of Methods [\textbf{NOM}], Number of Declared Methods [\textbf{NOM-A}] Number of Instance Variables [\textbf{NIV}] \textbf{File:} Total Lines [\textbf{F-TLOC}], Code Lines [\textbf{F-CLOC}], Blank Lines [\textbf{F-BLOC}], Statement Count [\textbf{F-STMT}], Declaration Statements [\textbf{F-DSTMT}], Executable Statements [\textbf{F-ESTMT}] \\ \midrule Documentation & \textbf{Method:} Comment Lines [\textbf{COMLOC}], Comment to Code Ratio [\textbf{CCR}], Comment Lines w/ Preceding Code [\textbf{CLWB}], Comment Ratio (Preceding) [\textbf{CCR-B}] \textbf{Class:} Comment Lines [\textbf{CCOM}], Comment to Code Ratio [\textbf{CCR-C}] \textbf{File:} Comment Lines [\textbf{F-COMLOC}], Comment-to-Code Ratio [\textbf{F-CCR}] \\ \midrule Coupling \& Inheritance & \textbf{Method:} Fan-In [\textbf{FI}], Fan-Out [\textbf{FO}], Component Reuse [\textbf{CR}] \textbf{Class:} Depth of Inheritance Tree [\textbf{DIT}], Base Classes [\textbf{BCs}], Derived Classes [\textbf{DCs}] \\ \midrule Python Specific & Maximum Indentation [\textbf{PMI}], Magic Numbers [\textbf{PMN}] \\ \midrule \multicolumn{2}{l}{\textbf{Statistical Metrics}} \\ \midrule Distribution & Entropy [\textbf{ENT}] \\ \midrule \multicolumn{2}{l}{\textbf{Process Metrics}} \\ \midrule Temporal & Age in Days [\textbf{AGE}], Bug Density [\textbf{BD}], Fix Commits [\textbf{FC}] \\ \midrule Churn-Based & Average Code Churn [\textbf{ACCH}], Maximum Code Churn [\textbf{MCCH}], Total Code Churn [\textbf{TCCH}], Total Modified Statements [\textbf{TMS}] \\ \midrule Commit-Based & Total Commits [\textbf{TC}], Commits Count [\textbf{CMC}], Maximum Commit LOC Changes [\textbf{MCLC}], Average Commit LOC Changes [\textbf{ACLC}], Total Commit Churn [\textbf{TCC}], Commit Churn Added [\textbf{CCA}], Commit Churn Deleted [\textbf{CCD}], Change Pattern Count [\textbf{CPC}] \\ \midrule Method-Based & Method Churn Added [\textbf{MCA}], Method Churn Deleted [\textbf{MCD}], Total Method Churn [\textbf{TMC}], Average Method LOC Changes [\textbf{AMLC}], Maximum Method LOC Changes [\textbf{MMLC}] \\ \midrule Developer-Based & Distinct Authors [\textbf{DA}], Average Developer Experience [\textbf{ADE}], Distinct Commit Names [\textbf{DCN}], Average Commits per Author [\textbf{ACA}], Average Code Churn per Author [\textbf{ACCA}] \\ \bottomrule \end{tabular}  \end{table*}

When no linked issue exists, the algorithm resorts to commit-level timing (line~16), which is the only remaining temporal information available in the absence of a linked issue. It retrieves the commit date and the repository's first stable release date. If the commit precedes the first stable release or the project has no stable release at all, the commit is classified as pre-release (line~17). 
This follows the standard assumption that post-release failures occur after the first official stable release. Moreover, the conservative nature of our heuristic ensures minimal risk of false post-release classifications.

We intentionally adopt a conservative policy and never classify a commit as post-release based solely on its timestamp. Due to non-linear release workflows, maintenance branches, and backports, timestamp-only evidence is unreliable for inferring post-release origin. In these cases, we prefer the neutral label \emph{unknown}.
If none of the above conditions apply, meaning the algorithm cannot derive reliable temporal or role-based evidence, the commit is left as \textit{unknown} (line~19).

To assess the reliability of this heuristic, we conducted a manual inspection of 400 randomly sampled software faults drawn from the collected commits. The sample was drawn uniformly across all projects, and each commit was manually labeled based on its complete issue and commit context. Manual annotation was performed by one author and cross-checked by a second author in ambiguous cases, until consensus was reached. Both authors involved in the process (a Ph.D. student and an Assistant Professor in Computer Engineering) have a solid background in software testing and fault classification.
Our manual analysis revealed that $\sim95\%$ of these cases were correctly classified as pre-release or post-release. The remaining instances were almost exclusively commits that were either not true bugs (e.g., documentation updates, refactorings) or lacked any associated issue or temporal information, and were therefore legitimately labeled as \textit{unknown} by the algorithm.

We remark that our goal with this heuristic is not to build a perfect classifier, but to obtain sufficiently reliable labels to support downstream modeling. The conservative design, combined with manual validation, suggests that residual mislabeling noise is limited and does not compromise the trends observed in our empirical analysis.

\subsection{Metric Collection}

A critical step in developing effective fault prediction models is selecting a comprehensive subset of software metrics that accurately characterizes residual and non-residual faults. 


We began by conducting a thorough review of existing literature to collect metrics previously demonstrated as effective predictors of software faults~\cite{caulo2020metrictaxonomy}. This resulted in a broad pool of 83 metrics, categorized into three main groups: product metrics, which describe static characteristics of the source code, statistical metrics, which capture the differences in the code variability, and process metrics, which capture historical and evolutionary aspects of the codebase (see Table~\ref{tab:metrics}).

The metrics adopted in this study cover a wide range of code characteristics at the method, class, and file levels. We collected them by using Scitools Understand \cite{scitools_understand} and ad-hoc scripts. \textit{Product metrics} describe static properties of the code and are grouped into complexity, size, documentation, coupling, inheritance, and Python-specific dimensions. Complexity indicators include Cyclomatic Complexity \texttt{[CC]}, Maximum Nesting Depth \texttt{[MND]}, and the Number of Paths \texttt{[NP]} at method level, with corresponding file-level variants such as \texttt{[F-CC]}, \texttt{[F-MND]}, and \texttt{[F-NPLOG]}. In addition, an extensive set of Halstead metrics (Difficulty, Length, Vocabulary, Volume, Effort, Maintainability Index, and operator/operand counts) captures the informational structure of the code and the cognitive effort required to understand it. Size metrics quantify dimensional properties at multiple granularities, including Lines of Code, declaration and executable statement counts, inputs and outputs, and the number of exits or early exits in methods. Class-level size measures (e.g., \texttt{[CLLOC]}, \texttt{[CDLOC]}, \texttt{[NOM]}) reflect structural regularities that influence maintainability. File-level size indicators complement these with global views of the code layout. Documentation-related metrics, including comment lines, comment-to-code ratios, and comments appearing before code, characterize the density and distribution of documentation across methods, classes, and files. Coupling and inheritance measures, such as Fan-In and Fan-Out at the method level, and the Depth of Inheritance Tree and the number of base or derived classes at the class level, provide information on dependency structure and the potential for fault propagation. Python-specific indicators, including maximum indentation and magic numbers, capture aspects that often correlate with readability and error-proneness.

\textit{Statistical metrics} complement them with a distribution-based measure of code regularity. Entropy \texttt{[ENT]} was computed using KenLM, trained on the ETH Py150 corpus of 150k Python files~\cite{raychev2016probabilisticmodel}. After training the language model, the entropy of each function in our dataset was computed, allowing us to estimate how typical or atypical a piece of code is relative to a large and diverse Python codebase.

\textit{Process metrics} describe the evolution of the code and how developers interact with it. Temporal indicators, such as Age in Days \texttt{[AGE]}, capture the maturity of a component. Churn-based measures quantify the extent of modifications performed over time, covering averages, maxima, and totals for both code churn and modified statements. Commit-based metrics capture the structure and magnitude of version control activity, including the total number of commits, commit-level line changes, added or deleted code, and repeated change patterns. Method-based measures capture fine-grained evolution through counts of method additions, deletions, and line-level modifications. Developer-based indicators quantify the social and organizational aspects of code evolution, such as the number of distinct authors, the distribution of contributions, and average developer experience, calculated as the average of all developers’ experience scores, where each score reflects how much, how long, and how recently a developer worked on the method. All product and process metrics were computed through purpose-built analysis scripts, validated across multiple runs to ensure consistency and correctness.

\begin{figure}[t] 
    \centering
    \includegraphics[width=1\columnwidth]{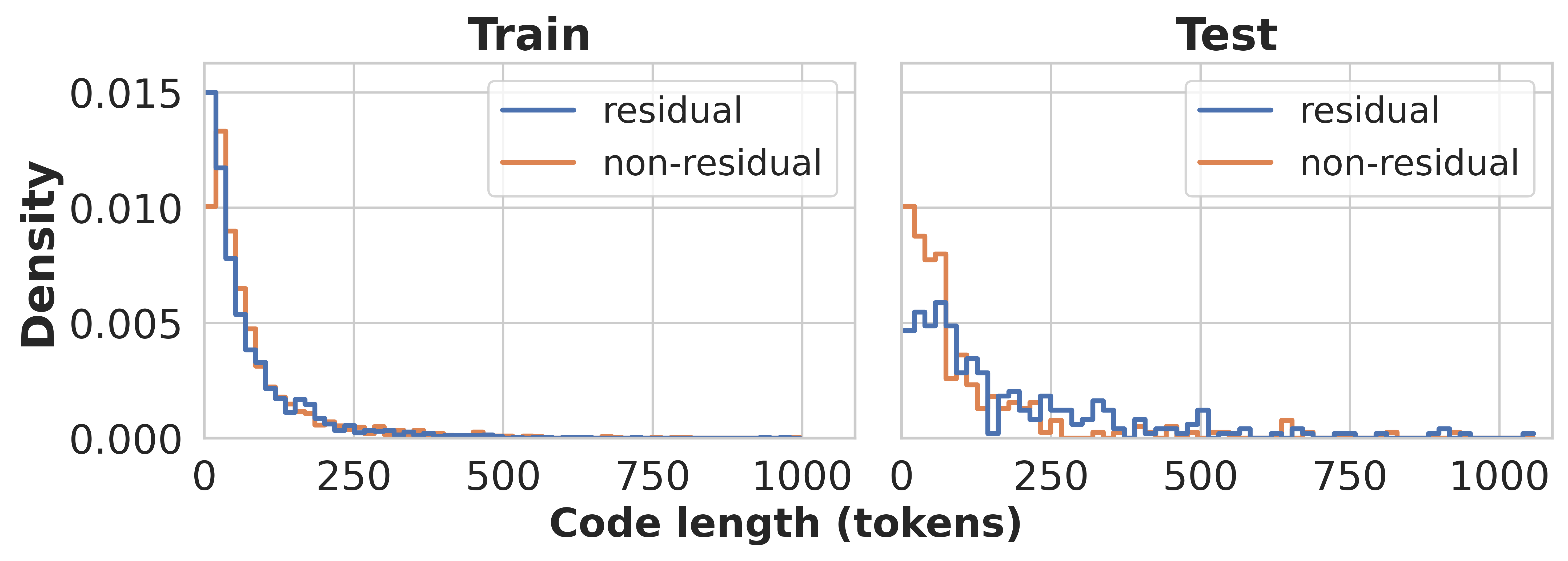} 
    
    \caption{Histograms of the train and test splits
showcasing the distribution of token counts across residual and non-residual faults}
    \label{fig:token_code_length}
    
\end{figure}

\section{Experimental Setup}
\label{sec:setup}
We conducted our experiments using both local development resources and large-scale computing infrastructure. Initial testing and model development were carried out on a standalone workstation equipped with a 12th Gen Intel Core i7-12700 processor (12 cores, 20 threads, up to 4.9 GHz), 32 GB of RAM, and an NVIDIA GeForce GTX 1660 SUPER GPU with 6 GB of dedicated memory. For the fine-tuning of deep learning models, we relied on NVIDIA A100 GPUs with 80 GB of memory; each training job was assigned 16 Intel Xeon CPU cores and 30 GB of RAM.


Building upon this computational setup, we prepared the data used for training and validation. We adopted a 90–10 random splitting strategy, allocating 90\% of the samples to the training set and 10\% to the validation set. The resulting dataset comprises 3570 faults in the training split (1770 residual and 1800 non-residual) and 450 in the validation split (213 residual and 217 non-residual). To evaluate generalizability beyond our primary dataset, we further employed BugsInPy~\cite{widyasari2020bugsinpy}, a benchmark of verified Python bugs. After classifying each bug, we obtained 284 residual and 221 non-residual faults. All splits were checked to ensure that no commit appears in more than one partition, preventing data leakage. In total, the final corpus contains 4525 faults, of which 2267 are residual and 2238 non-residual. The 90/10 split models a single-ecosystem scenario in which an organization trains the models on historical data and tests them on real-world, previously unforeseen use case scenarios. BugsInPy serves as a cross-project evaluation benchmark, as it contains entirely separate commits and code with no overlap with the training corpus. Specifically, the training corpus comprises $357$ different projects, while BugsInPy contains $16$ different projects. Out of these $16$, $10$ projects are also in the training corpus.


\begin{table}[t]
\caption{Statement-level statistics by split and residual label. 
}
\label{tab:stats_split_residual}
\centering
\small
\begin{tabular}{
>{\arraybackslash}m{2.5cm} | 
>{\arraybackslash}m{1.5cm}
>{\arraybackslash}m{1.5cm}
>{\arraybackslash}m{1,5cm}
}
\toprule
\textbf{Group} & 
\textbf{Unique Statements} & 
\textbf{Unique Tokens} & 
\textbf{Average Tokens} \\
\midrule
train (residual)   & 29{,}459 & 40{,}710 & 3.34 \\
train (non-residual)   & 29{,}992 & 44{,}880 & 3.69 \\ 
test (residual)        & 9{,}470  & 13{,}808 & 4.19 \\
test (non-residual)        & 4{,}612  & 7{,}900  & 3.96 \\
\bottomrule
\end{tabular}

\end{table}

\figurename{}~\ref{fig:token_code_length} provides an initial overview of the dataset by illustrating the distribution of code lengths, expressed in number of tokens, across residual and non-residual faults. Several trends emerge. First, code length in both training and test splits is right-skewed: while most samples are relatively short, a long tail of larger snippets appears in both classes. Second, in the training split, residual and non-residual distributions nearly overlap, with most examples below 250 tokens and only a minority exceeding 500. In contrast, the test split shows a slightly higher proportion of longer residual examples, reflecting the increased complexity of the faults in BugsInPy.

To complement these observations, \tableautorefname{}~\ref{tab:stats_split_residual} reports statement- and token-level statistics. In the training set, residual and non-residual faults contain a comparable number of unique statements (29k–30k), but non-residual examples include more unique tokens overall, leading to a higher average token count per statement. This distinction becomes even more pronounced in the test set, where residual cases display a higher average token density (4.19 vs. 3.96). Taken together, these findings indicate that residual bugs often appear in structurally dense code, while also confirming that the two classes remain well balanced in their structural properties.


\section{Experimental Results}
\label{sec:evaluation}
In this section, we present the results of our experimental campaign on predicting \emph{residual faults} in Python systems. Detecting residual faults is intrinsically
challenging: these bugs often differ only subtly from pre-release defects, and no straightforward lexical or syntactic patterns clearly separate the two. Because of this, we investigate the task through a progressively more informed sequence of analyses, moving from general-purpose reasoning models to domain-adapted models and finally to metric-based learners.

To guide our evaluation, we define four research questions:

\begin{itemize}
    \item \textbf{RQ1:} \textit{Can code representations learned by LLMs discriminate residual from non-residual faults?}
    \item \textbf{RQ2:} \textit{Can software metrics improve residual fault prediction?}
    \item \textbf{RQ3:} \textit{Which software metrics contribute most to the classification?}
    \item \textbf{RQ4:} \textit{Do software metrics and LLM-based code representations encode the same or complementary information?}
\end{itemize}

The order of our analyses follows a progression.  We begin by assessing whether code representations learned by Large Language Models, including both proprietary systems and fine-tuned open-weight models, can capture patterns that distinguish residual from non-residual faults (RQ1). 
We then shift to an orthogonal perspective and examine whether traditional software metrics, such as those used in anomaly detection or supervised classification, provide a clearer signal for this prediction task (each model using the same fixed system prompt, which instructed decisions of the supervised models, leveraging explainability techniques to characterize their contribution (RQ3). Finally, we investigate whether software metrics and code-based representations encode overlapping or complementary information, and thus whether combining them can improve residual-fault prediction (RQ4).

\subsection{RQ1: Can code representations learned by LLMs discriminate residual from non-residual faults?}

In this RQ, we investigate whether the internal representations of Large Language Models (LLMs) can distinguish between residual and non-residual faults. We divide our analysis into two parts: state-of-the-art closed-source models prompted in a zero-shot setting and open-weight models fine-tuned specifically for this task.

\subsubsection{Closed-Source General-Purpose LLMs}

We first evaluated three state-of-the-art proprietary models, namely Gemini 3 Pro, Claude 4.5 Sonnet, and GPT-4.1, on our test set as a baseline, using them through their respective APIs.
Before running the evaluation, we verified that all three models understood the concept of a residual fault. When asked directly, each model correctly described it as a bug that escaped the testing phase, confirming that a lack of knowledge did not hinder the task itself. To ensure comparability, we evaluated each model using the same fixed system prompt, which defined a residual fault and asked them to classify the input code.

Their performance is summarized in Table~\ref{tab:closed_models_results}. The three models exhibit two main approaches that underscore the difficulty of the task and the absence of simple lexical or structural cues that these models can reliably exploit.

Gemini 3 Pro achieves the highest F1-score (0.72); however, its behavior is dominated by an extremely aggressive classification strategy. Out of the 284 residual faults in the test set, Gemini correctly classifies 280 as post-release (TP) and misses only 4 (FN), yielding a recall of 0.99 on this class. However, this comes at the cost of misclassifying almost every non-residual instance: among the 221 pre-release faults, it correctly flags only 10 as pre-release (TN) and incorrectly labels the remaining 211 as post-release (FP). This results in a misleadingly strong F1, driven not by discriminative ability but by systematically favoring the residual label. In practice, Gemini 3 Pro behaves as if “residual” were the default prediction.

\begin{table}[t]
\caption{Performance of closed-source LLMs on the residual fault prediction task with 95\% CI.}
\label{tab:closed_models_results}
\centering
\resizebox{\columnwidth}{!}{%
\begin{tabular}{lcccc}
\toprule
\textbf{Model} & \textbf{Accuracy} & \textbf{Precision} & \textbf{Recall} & \textbf{F1} \\
\midrule
Gemini~3~Pro      & $\mathbf{0.574}_{\scriptscriptstyle[.53,.62]}$ & $0.570_{\scriptscriptstyle[.53,.61]}$ & $\mathbf{0.986}_{\scriptscriptstyle[.97,1.0]}$ & $\mathbf{0.723}_{\scriptscriptstyle[.69,.76]}$ \\
Claude~4.5~Sonnet & $0.535_{\scriptscriptstyle[.49,.58]}$ & $\mathbf{0.615}_{\scriptscriptstyle[.55,.68]}$ & $0.461_{\scriptscriptstyle[.40,.52]}$ & $0.527_{\scriptscriptstyle[.47,.58]}$ \\
GPT--4.1          & $0.461_{\scriptscriptstyle[.42,.50]}$ & $0.531_{\scriptscriptstyle[.46,.60]}$ & $0.359_{\scriptscriptstyle[.30,.42]}$ & $0.429_{\scriptscriptstyle[.37,.48]}$ \\
\bottomrule
\end{tabular}}
\end{table}

Claude 4.5 Sonnet adopts a much more conservative stance. It produces predictions for 416 out of 505 test instances, abstaining on 89 cases. On the subset where it does emit a label, Claude correctly identifies 132 pre-release faults (TN) and produces 62 false positives, showing a noticeably better ability than Gemini to recognize non-residual faults. At the same time, it correctly classifies only 84 residual faults (TP) and misclassifies 138 as pre-release (FN), yielding a recall of 0.38 on the residual class. This pattern indicates a reluctance to assign the residual label: Claude tends to default to pre-release unless the signal is overwhelmingly clear, which reduces false alarms but systematically misses many true residual faults.

GPT-4.1 falls in the same trap, remaining ineffective overall. Like Claude, its confusion matrix still shows a strong difficulty in capturing residual faults. Among the 221 pre-release cases, GPT-4.1 correctly identifies 131 (TN) and mislabels 90 as post-release (FP), achieving a precision comparable to Claude on the residual class. However, on the 284 residual cases it correctly classifies only 102 (TP) and misclassifies 182 as pre-release (FN), yielding a recall of 0.36 and an F1 of 0.43. Rather than favoring one class as strongly as Gemini, GPT-4.1 distributes errors across both, suggesting uncertainty. To assess whether the observed performance differences are statistically significant, we apply McNemar's test~\cite{mcnemar1947note}, a non-parametric test designed to compare two classifiers evaluated on the same set of instances. The test operates on the $2 \times 2$ contingency table of discordant predictions, i.e., cases where one model is correct, and the other is not, and is the standard choice when per-sample predictions are available from a single test set~\cite{dietterich1998approximate}. Results show that GPT-4.1 differs significantly from both Gemini~3~Pro ($p{=}0.002$) and Claude~4.5~Sonnet ($p{=}0.019$), whereas Gemini and Claude do not differ significantly ($p{=}0.260$), indicating that, despite opposite classification biases, their overall error rates are comparable.

\begin{mybox}{Closed Source LLMs on Code}
Closed-source LLMs fail to predict residual faults reliably.
Gemini~3~Pro inflates its F1 by labeling almost everything as residual, while Claude~4.5~Sonnet and GPT--4.1 rarely detect residual faults at all.
None of the models learns a meaningful boundary, confirming that general-purpose LLMs are ineffective for this task. 
\end{mybox}

\subsubsection{Fine-Tuned Deep Learning Models}

We next evaluate a set of trained deep learning models on the same prediction task. The goal here is twofold: (i) to assess whether fine-tuning on code can overcome the limitations observed with closed-source LLMs, and (ii) to determine whether code representations extracted by large pretrained models contain signals that generalize beyond project-specific context. For this purpose, we selected three representative architectures that cover the state-of-the-art DL models: \emph{CodeT5+ for Classification}~\cite{wang2023codet5plus} (770M parameters), an encoder–decoder model optimized for code understanding; \emph{DeepSeek-Coder}~\cite{wang2023deepseek} (1.3B parameters), a transformer decoder pretrained specifically on code repositories; and \emph{CodeLlama}~\cite{rozière2024codellamaopenfoundation} (7B parameters), a decoder-only backbone specialized for code generation.
Before training, each snippet was processed through our normalization pipeline, which removes comments and docstrings, anonymizes identifiers and literals, and cleans possible noise around the method (e.g., wrongly extracted keywords). This prevents models from relying on project-specific tokens, forcing them instead to use structural and semantic cues.

\begin{table}[t]
\caption{Performance of trained deep learning models on the residual fault prediction task
with 95\% CI.}
\label{tab:dl_models_results}
\centering
\resizebox{\columnwidth}{!}{%
\begin{tabular}{lcccc}
\toprule
\textbf{Model} &
\textbf{Accuracy} &
\textbf{Precision} &
\textbf{Recall} &
\textbf{F1} \\
\midrule
CodeLlama       & $\mathbf{0.438}_{\scriptscriptstyle[.40,.48]}$ & $\mathbf{0.500}_{\scriptscriptstyle[.43,.57]}$ & $0.327_{\scriptscriptstyle[.27,.38]}$ & $0.396_{\scriptscriptstyle[.34,.45]}$ \\
CodeT5+         & $0.406_{\scriptscriptstyle[.36,.45]}$ & $0.470_{\scriptscriptstyle[.41,.53]}$ & $\mathbf{0.440}_{\scriptscriptstyle[.38,.50]}$ & $\mathbf{0.455}_{\scriptscriptstyle[.40,.51]}$ \\
DeepSeek-Coder  & $0.388_{\scriptscriptstyle[.35,.43]}$ & $0.426_{\scriptscriptstyle[.35,.50]}$ & $0.254_{\scriptscriptstyle[.20,.30]}$ & $0.318_{\scriptscriptstyle[.26,.37]}$ \\
\bottomrule
\end{tabular}}
\end{table}

Despite their architectural differences, the three models exhibit limited predictive ability (Table~\ref{tab:dl_models_results}). CodeT5+ achieves the highest F1 (0.46), followed by CodeLlama (0.40) and DeepSeek-Coder (0.32). However, none of the models reaches performance levels comparable to supervised metric-based approaches.

The confusion matrices reveal distinct behavioral profiles. CodeT5+ detects the most residual faults, correctly identifying 125 true positives, but at the cost of 141 false positives and still missing 159 residual cases (FN). CodeLlama trades recall for precision: it identifies 93 true positives with 93 false positives, yielding the highest accuracy (0.44) and precision (0.50) among the three, but missing 191 residual faults. DeepSeek-Coder performs worst overall, detecting only 72 residual faults while misclassifying 97 non-residual cases and leaving 212 residual faults undetected. McNemar's test confirms that CodeLlama and CodeT5+ do not differ significantly ($p{=}0.202$), nor do CodeT5+ and DeepSeek-Coder ($p{=}0.460$), whereas CodeLlama significantly outperforms DeepSeek-Coder ($p < 0.001$).

\begin{mybox}{Fine-tuned LLMs on Code}
Fine-tuned deep learning models also struggle to predict residual faults.
CodeT5+ achieves the best F1 (0.46) but at the cost of high false positives, while CodeLlama and DeepSeek-Coder miss the majority of residual cases.
Despite architectural differences, none of the models extracts robust signals from code, confirming that residual fault prediction requires more specialized representations.
\end{mybox}

\subsection{RQ2: Can software metrics improve residual fault prediction?}

In this RQ, we investigate whether software metrics can help predict residual faults. To do so, we explicitly compare unsupervised anomaly detection algorithms with supervised classification models. All features were standardized to a zero mean and unit variance for both experimental settings.

\subsubsection{Unsupervised Anomaly Detection}

Unsupervised methods are appealing in this context because residual faults are relatively rare. For this setting, we use Isolation Forest (\emph{IF})\cite{liu2008isolation}, One-Class SVM (\emph{OC-SVM} \cite{scholkopf2001estimating}, and Local Outlier Factor (\emph{LOF}) \cite{breunig2000lof}. They model expected behavior and flag statistically atypical cases, helping identify potential defects or quality anomalies \cite{ding2020improved, moussa2024oneclasssupport, rabih2025highly}.

\begin{table}[t]
\caption{Performance of unsupervised models
with 95\% CI.}
\label{tab:metrics_unsupervised_results}
\centering
\resizebox{\columnwidth}{!}{%
\begin{tabular}{lcccc}
\toprule
\textbf{Model} &
\textbf{Accuracy} &
\textbf{Precision} &
\textbf{Recall} &
\textbf{F1} \\
\midrule
IsolationForest    & $\mathbf{0.523}_{\scriptscriptstyle[.48,.57]}$ & $0.642_{\scriptscriptstyle[.57,.72]}$ & $\mathbf{0.342}_{\scriptscriptstyle[.29,.40]}$ & $\mathbf{0.446}_{\scriptscriptstyle[.39,.50]}$ \\
OneClassSVM        & $0.497_{\scriptscriptstyle[.45,.54]}$ & $0.650_{\scriptscriptstyle[.55,.74]}$ & $0.229_{\scriptscriptstyle[.18,.28]}$ & $0.339_{\scriptscriptstyle[.28,.40]}$ \\
LocalOutlierFactor & $0.493_{\scriptscriptstyle[.45,.54]}$ & $\mathbf{0.818}_{\scriptscriptstyle[.70,.93]}$ & $0.127_{\scriptscriptstyle[.09,.17]}$ & $0.220_{\scriptscriptstyle[.16,.28]}$ \\
\bottomrule
\end{tabular}}
\end{table}

As shown in Table~\ref{tab:metrics_unsupervised_results}, unsupervised anomaly detectors perform poorly on this task. Although they achieve reasonable precision on the residual class, they detect only a small portion of the true residual faults. IF correctly identifies 97 residual cases (TP) but misses 187 (FN), while OC-SVM and LOF degrade further to 64 and 35 true positives, respectively, leaving more than two hundred residual faults undetected (FN 220 and 249, respectively). In all three methods, false positives remain modest, but this stems from a consistent pattern: they overwhelmingly classify faults as non-residual, rarely flagging anomalies. McNemar's test reveals no significant pairwise difference among the three detectors ($p \geq 0.13$ for all pairs), confirming that their poor performance is a structural limitation of the anomaly-detection paradigm for this task. These results confirm that residual faults do not manifest as outliers in metric space, and that unsupervised anomaly detectors are fundamentally unsuited for this prediction task.

\begin{mybox}{Unsupervised ML Models on Software Metrics}

Unsupervised anomaly detectors (IF, OC-SVM, LOF) fail to identify residual faults reliably.
While they maintain reasonable precision, they suffer from extremely low recall, missing the vast majority of residual cases.
Residual faults do not manifest as clear statistical outliers in the software metric space, rendering anomaly detection ineffective for this task.

\end{mybox}

\subsubsection{Supervised Classification}

\begin{table}[t]
\caption{Performance of supervised models
with 95\% CI.}
\label{tab:metrics_models_results}
\centering
\resizebox{\columnwidth}{!}{%
\begin{tabular}{lcccc}
\toprule
\textbf{Model} &
\textbf{Accuracy} &
\textbf{Precision} &
\textbf{Recall} &
\textbf{F1} \\
\midrule
RandomForest & $\mathbf{0.636}_{\scriptscriptstyle[.59,.68]}$ & $\mathbf{0.622}_{\scriptscriptstyle[.57,.67]}$ & $\mathbf{0.898}_{\scriptscriptstyle[.86,.93]}$ & $\mathbf{0.735}_{\scriptscriptstyle[.70,.77]}$ \\
XGBoost      & $0.610_{\scriptscriptstyle[.57,.65]}$ & $0.611_{\scriptscriptstyle[.56,.66]}$ & $0.845_{\scriptscriptstyle[.80,.89]}$ & $0.709_{\scriptscriptstyle[.67,.75]}$ \\
CatBoost     & $0.628_{\scriptscriptstyle[.59,.67]}$ & $\mathbf{0.622}_{\scriptscriptstyle[.57,.67]}$ & $0.870_{\scriptscriptstyle[.83,.91]}$ & $0.724_{\scriptscriptstyle[.68,.76]}$ \\
\bottomrule
\end{tabular}}
\end{table}

Supervised models trained on the same metrics perform significantly better and reveal a different set of trade-offs (Table~\ref{tab:metrics_models_results}). RandomForest, XGBoost, and CatBoost reach F1-scores between 0.71 and 0.73 and show substantially improved sensitivity to residual faults. RandomForest, for example, correctly identifies 255 residual faults (TP: 255, FN: 29) but also produces 155 false positives (TN: 66). XGBoost and CatBoost behave similarly, obtaining 241–248 true positives while producing roughly 150 false positives each. Compared to the unsupervised baselines, these supervised learners recover more than three times as many residual faults and reduce false negatives, demonstrating that software metrics contain meaningful predictive signals. McNemar's test shows no significant difference among the three supervised learners ($p \geq 0.11$ for all pairs), indicating that the predictive signal resides in the metrics themselves rather than in the choice of classifier. In contrast, cross-family comparisons confirm that supervised models significantly outperform both unsupervised detectors (RandomForest vs.\ IsolationForest, $p < 0.001$) and closed-source LLMs (RandomForest vs.\ Gemini~3~Pro, $p = 0.003$), despite the latter's nominally comparable F1.

When contrasted with the other deep learning models studied, the supervised ones occupy an interesting middle ground. They do not match Gemini~3~Pro's extreme recall (which comes from labeling nearly everything as residual). Still, unlike Gemini, their high TP counts arise from actual discriminative patterns rather than a degenerate decision rule. At the same time, they outperform Claude~4.5~Sonnet and GPT-4.1, CodeLlama, CodeT5+, and DeepSeek-Coder in both TP and FN ratios, revealing a stronger ability to separate residual from non-residual cases than any model relying solely on representations learned from code. Yet their high false-positive rates reveal the limits of metric-only features.

\begin{mybox}{Supervised ML Models on Software Metrics}

Supervised learners trained on software metrics significantly outperform unsupervised detectors and many deep learning models.
RandomForest, XGBoost, and CatBoost correctly identify 241–255 residual cases, reducing false negatives by an order of magnitude.
However, all three models still generate several false positives: metric-based supervision captures useful predictive signals but struggles to achieve a clean separation between pre-release and post-release faults, positioning these models as safeguards rather than definitive tools.

\end{mybox}

\subsection{RQ3: Which software metrics contribute most to the classification?}

\begin{figure}[t]
\centering
\includegraphics[width=1\columnwidth]{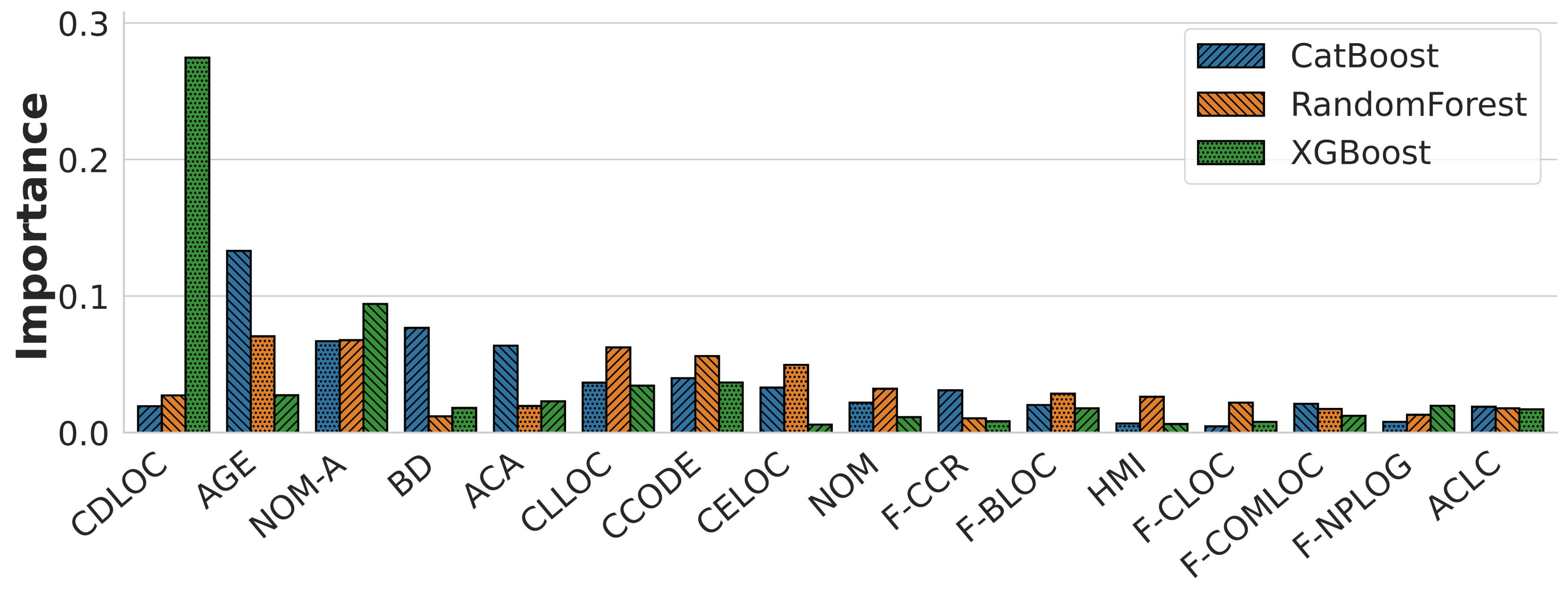}
\caption{Top-10 feature importances for the three supervised models used in RQ2}
\label{fig:features_all}
\end{figure}

To identify the properties that distinguish residual from non-residual faults, we examine the feature-importance rankings produced by the three supervised learners from RQ2 and complement them with a SHAP-based directional analysis~\cite{lundberg2017unified}. \figureautorefname{}~\ref{fig:features_all} reports the top-10 metrics for each model.
The figure reports each model’s feature-importance scores, i.e., the model-specific estimates of how much each metric contributes to prediction, normalized so that importances sum to 1.

Across all models, it emerges that process metrics play a central role, in particular the age of the method (\emph{AGE}), which is highly ranked in all three learners. The SHAP analysis indicates that higher AGE pushes the prediction towards the residual-fault class, suggesting that long-lived components are more likely to host faults that escape pre-release detection. In contrast, bug density (\emph{BD}) and average commits per author (\emph{ACA}) have an opposite effect: their SHAP means are negative, meaning that larger BD and ACA values are associated with a higher probability of non-residual faults. As expected, modules with many historical defects and intense patching activities have a reduced chance that further faults remain hidden.

Structural and size-oriented class metrics are equally prominent. \emph{CDLOC} (class declaration lines), \emph{NOM-A} (number of declared methods), \emph{CLLOC} (class LOC), \emph{CCODE}, \emph{CELOC}, and \emph{NOM} all appear in the top-10 lists, especially for XGBoost and RandomForest. For these metrics the SHAP means are consistently positive, i.e., larger classes increase the likelihood of a residual-fault prediction. 

File-level size and navigability measures provide additional explanatory power. \emph{F-BLOC} (blank lines) and \emph{F-NPLOG} (logarithm of paths) are important for XGBoost and, according to SHAP, higher values of both push towards the residual-fault class, indicating that large and structurally complex files are risk-prone. \emph{F-COMLOC} (file comment lines) is relevant for CatBoost and also shows a positive SHAP mean, suggesting that highly documented components are not necessarily less risky. Conversely, \emph{F-CCR} (comment-to-code ratio) tends to have a small but negative SHAP mean in CatBoost, showing that well-documented and written code is less prone to host residual faults.

While the three learners share these general tendencies, their emphases differ. XGBoost focuses on a small set of structural metrics (CDLOC, NOM-A, CLLOC, F-BLOC, F-NPLOG), all of which exhibit a clear positive SHAP direction. CatBoost distributes importance more evenly between process and structural dimensions, with AGE, ACA and BD among the most influential process metrics. RandomForest assigns similar relative importance to AGE, NOM-A, CLLOC, CDLOC, CCODE, and CELOC, but its SHAP means are close to zero, indicating weaker directional effects. 

\begin{mybox}{Key Software Metrics for Fault Prediction}
Residual-fault prediction is driven by a combination of historical process indicators and specific structural size characteristics, with AGE, class/file size, and dependencies consistently associated with an increased residual-fault proneness. Post-release faults in Python mostly arise in older, larger, structurally complex, and badly documented components.
\end{mybox}

\subsection{RQ4: Do software metrics and LLM-based code representations encode the same or complementary information?}

Our results show that software metrics prove more useful than code representations for cross-project prediction tasks. However, this does not imply that one representation is inherently superior. Our question is whether code can provide an additional, orthogonal source of information that can be combined with metrics to improve prediction when conditions allow. To explore this, we compared traditional software metrics with dense code representations extracted from the pretrained CodeLlama model. Embeddings were obtained using its tokenizer and the penultimate transformer layer, followed by mean pooling to produce a single 4,096-dimensional vector per function. The model weights remained frozen, ensuring that the embeddings capture syntactic and semantic patterns learned during pretraining.

Using this representation, we examined the relationship between metrics and embeddings. After standardization, both spaces were reduced with PCA~\cite{abdi2010principal}, yielding 29 components for metrics and 28 for embeddings, each preserving more than 95\% of the variance. The association between the two is limited: the maximum absolute Spearman correlation between any metric and embedding component is 0.27, and the average maximum correlation per component is roughly 0.11–0.12. Canonical Correlation Analysis (CCA)~\cite{hotelling1992relations} confirms this observation. Although the first few canonical correlations are moderate (about 0.71, 0.53, and 0.50), the mean across the first twenty is only around 0.24, indicating that the shared subspace is small.
The geometric distinction between the two representations is evident in the \figureautorefname{}~\ref{fig:pca-orthogonality}, where components derived from metrics and embeddings form clearly separated clusters with almost orthogonal regression directions.

Taken together, these results suggest that improving testing does not require replacing metrics with embeddings or vice versa, but combining them. Metrics could help identify where additional testing and review should be directed, while embeddings capture semantic aspects of faults. 

\begin{mybox}{Differences between Code and Metrics}
Metrics and embeddings share little information. After PCA, their components reach a
maximum Spearman correlation of 0.27, typically around 0.11–0.12. CCA confirms this: the mean of the first twenty components is $\approx 0.24$.
Their PCA clusters are nearly orthogonal, indicating that metrics capture structural
factors while embeddings provide complementary semantic cues.
\end{mybox}

\begin{figure}[t]
    \centering
    \includegraphics[width=0.8\columnwidth]{images_evaluation_regression_pca.pdf}
    \caption{PCA projection of metric and embedding components. The two X’s represent the centroids. The two clusters exhibit a near-orthogonality of the two representations.}
    \label{fig:pca-orthogonality}
\end{figure}

\section{Discussion}
\label{sec:discussion}
Our results highlight a clear separation between code-based and metric-based approaches to residual fault prediction. Deep learning models such as CodeT5+, DeepSeek-Coder, and CodeLlama struggle to form a reliable decision boundary and consistently miss true residual faults. They also require substantial resources: training CodeT5+ and DeepSeek-Coder on an NVIDIA A100 with 80 GB of memory took about 1.3 hours, and CodeLlama took about 3 hours, with inference remaining slow and costly.

In contrast, traditional supervised models trained on software metrics are lightweight, fast, and more effective. On a standard server with a GTX 1660 Ti and an Intel Core i7-12700, they train in minutes and achieve higher recall. The precision of these models remains moderate, but prior work has shown that ~\cite{herbold2019costs, tunkel2022exploring} what matters in practice is the trade-off between inspection effort saved and residual faults caught. Our models retain $0.85-0.9$ recall while substantially reducing the number of commits requiring inspection, effectively acting as safety filters that preserve most escaping defects at a manageable additional validation cost
Their interpretability further reveals the structural and historical patterns associated with fault survival, including age, size, churn, and complex modification histories. McNemar's test shows no significant difference among RandomForest, XGBoost, and CatBoost ($p \geq 0.11$), indicating that the predictive power resides in the metrics rather than in a particular learning algorithm.

Metrics alone, however, do not capture all relevant information. Our representational analysis shows that metrics and embeddings encode largely independent signals: the maximum correlation between PCA components is only 0.27, typical values are around 0.11 to 0.12, and the mean canonical correlation across the first twenty components is about 0.24. The PCA projection forms two almost orthogonal clusters, indicating that embeddings carry semantic cues that metrics do not, even if those cues are insufficient on their own. These results point to a clear direction for future work. Models that combine structural indicators from metrics with semantic information from embeddings may achieve better stability and cross-project generalization. We conducted preliminary experiments with hybrid models that concatenate software metrics and code embeddings as input features. However, these models consistently produced degenerate behavior: when relying on metrics alone, performance remained stable, but whenever the models attempted to leverage the code representation, predictions degraded. We omitted detailed results for brevity, but the observed pattern suggests that the high dimensionality and variability of code embeddings overwhelm the compact metric signal during joint training, a challenge that warrants dedicated investigation with larger and more diverse datasets.

There remains a central question in our study: how can we improve the testing of residual faults? Our results suggest a two-part answer. First, testing should prioritize the components that metrics identify as high risk, such as old, large, highly modified, and frequently touched code, because these structural and historical factors strongly correlate with fault survival. Second, testing should incorporate semantic cues, since metrics and embeddings capture complementary information. Combining structural indicators with semantic patterns offers a more complete view of where testing is most likely to miss defects that later manifest in production. 
Our statistical analysis supports this direction: the stability of supervised performance across three independent classifiers suggests that the metric signal is robust, while the orthogonality between metrics and embeddings ($\rho_{\max}{=}0.27$) indicates substantial room for improvement through hybrid representations. Modern LLMs are well-positioned to provide such semantic cues~\cite{mockus2025code, mockus2025leveraging, abreu2025moving}: they have proved to hold great results when specifically fine-tuned on the same projects under test, but the challenge lies in capturing the extreme variability that the code representations have in cross-project scenarios.

\section{Related Work}
\label{sec:related}

Software Fault Prediction (SFP) has long been a central research area in software engineering, aiming to identify defect-prone components early in the development lifecycle to guide testing and quality assurance. Foundational studies demonstrated the predictive power of static code metrics, focusing on size~\cite{Akiyama1971}, control-flow complexity~\cite{McCabe1976}, and cognitive complexity~\cite{Halstead1977}. These early works laid the groundwork for leveraging structural indicators to infer software quality.

As programming paradigms evolved, object-oriented metrics were introduced to capture higher-level design properties such as coupling, cohesion, and inheritance~\cite{Chidamber1994}. Empirical evidence confirmed the effectiveness of these metrics in predicting faults across a wide range of systems~\cite{Briand2000}. 
In parallel, process metrics, including code churn~\cite{Graves2000}, change history~\cite{Ohlsson1996}, and developer activity~\cite{Hassan2009}, were adopted to reflect the dynamic aspects of software development. These metrics enhanced prediction models by incorporating historical and human-centric signals, often improving robustness~\cite{Radjenovic2013}.

Several systematic reviews have chronicled the evolution of SFP approaches. Catal et al.~\cite{Catal2009} emphasized the growing shift toward machine learning (ML)-based models and highlighted the need for publicly available datasets like NASA MDP~\cite{NASA_MDP} and PROMISE~\cite{PROMISE2005}. A follow-up study~\cite{Catal2011} further reported that method-level metrics and ML techniques were gaining traction due to their improved granularity and adaptability.

The emergence of DL has transformed the SFP landscape by enabling models to learn directly from raw source code. Akimova et al.~\cite{Akimova2021} reviewed DL-based SFP methods, noting the growing relevance of attention mechanisms and transformer-based architectures. Qiao et al.~\cite{Qiao2020} showed that DL models outperform traditional approaches by capturing syntactic and semantic patterns, while Feng et al.~\cite{Feng2020} demonstrated the utility of pretrained models like CodeBERT in learning long-range code dependencies. However, these models are opaque in their decision-making and demand significant computational resources, which may limit their adoption in production.

Alternative paradigms have also been explored to overcome the limitations of static models. Dynamic analysis techniques, such as those presented by Zhang et al.~\cite{Zhang2013}, observe runtime behavior to detect execution-specific faults. Natural Language Processing (NLP) methods mine commit logs and bug reports for fault signals~\cite{Hata2012}, and graph-based approaches model control-flow and data-flow dependencies using program graphs~\cite{Li2017}. While effective in capturing rich structural and behavioral information, these methods often suffer from scalability issues and reduced transparency, making them less accessible for practical deployment in industry settings.

A persistent limitation in the SFP literature is the lack of focus on residual (post-release) faults. Most studies treat all defects as temporally undifferentiated, even when logs or timestamps are available~\cite{Wang2020TSE,DallaPalma2022TSE}. Cross-project studies~\cite{Li2021,Liu2025,Lomio2022EMSE} similarly assume a homogeneous fault distribution, ignoring the operational and contextual challenges associated with post-release failures. This modeling assumption restricts the applicability of such models in production, where residual faults can be the most costly and disruptive~\cite{rwemalika2019industrialstudy}.

Recent work at Meta~\cite{mockus2025code, mockus2025leveraging, abreu2025moving} demonstrates that risk models built on prior defects can effectively identify risky changes within a single organization, reducing outages. In that setting, models are trained and deployed on the same codebase with stable distributions of developers, processes, and dependencies. Our evaluation setting differs precisely in this regard: we train on one dataset and test on BugsInPy, an independent benchmark comprising different projects, developers, and histories. Under such cross-project distribution shifts, code-level representations become unreliable because structural idioms vary widely across repositories. Process and product metrics, by contrast, provide a compact and project-agnostic representation of development dynamics.

Our work targets residual fault prediction, explicitly focusing on predicting residual faults. It contrasts interpretable software metrics with code-based features and evaluates them in a cross-project setting. Our methodology emphasizes model transparency through metric-level interpretability, achieving strong predictive performance while providing new insights into the most effective features and models for post-release reliability assessment. To our knowledge, no prior work has isolated residual fault prediction as a distinct task and evaluated it across projects using multiple models.


\section{Threats to validity}
\label{sec:threats}

\noindent
\textbf{Language Scope.}
Our study focuses exclusively on Python systems, selected for their ubiquity in AI, scripting, and rapid-development ecosystems~\cite{ray2014large, beller2016oops, pradel2018deepbugs}. Python’s dynamic typing, and runtime-dependent behavior may increase the likelihood of defects surfacing only after deployment, potentially amplifying the prevalence and characteristics of residual faults. As a result, caution is required when generalizing our findings to statically typed or lower-level languages such as Java or C++, where type systems, compilation pipelines, and memory-safety constraints shape different fault distributions and testing regimes. Future work will extend our methodology to multi-language datasets to assess whether the patterns observed for Python residual faults hold across programming paradigms.



\noindent
\textbf{Classification Heuristic Bias} Our residual vs. non-residual labeling relies on a heuristic that combines temporal evidence, issue metadata, and reporter identity. While this approach was deliberately designed to be conservative, favoring the unknown label when evidence is insufficient, it may introduce bias in the distribution of labels. In particular, commits without an associated issue are only classified as pre-release if they predate the first stable release, whereas no commit is labeled as post-release based solely on timestamp information. This choice avoids systematic misclassification due to non-linear release workflows, maintenance branches, or backports, but may underrepresent post-release faults in projects with sparse or inconsistent issue tracking. Although our manual validation demonstrates high accuracy ($\sim95\%$) on a random sample, some residual noise may still be present in the automatically derived labels. We mitigate this risk by (i) publicly releasing the full labeling code and feature definitions, enabling inspection and replication, and (ii) conducting sensitivity analyses confirming that the main experimental trends persist even when commits labeled as unknown are excluded.

\noindent
\textbf{Model Coverage.}
We evaluated a representative selection of traditional ML models, fine-tuned deep learning models, and prompt-based LLMs. While these choices span the current spectrum of fault prediction techniques, each model has inherent architectural biases and differing capacities for code representation. Our goal was not to identify a universally superior model, but to investigate whether diverse paradigms can effectively leverage software metrics and code representations for residual fault prediction.

\noindent
\textbf{Software Metric Selection.}
Our metric set combines product, process, statistical, and Python-specific features drawn from prior defect-prediction literature. Excluding highly sparse or trivially correlated metrics may further influence model behavior. To confirm this, we conducted ablation analyses, confirming that the main performance trends persist after removing the redundant metrics~\cite{yang2018ridge, jiarpakdee2016redundant, jiarpakdee2020impact}. 


\section{Conclusion and Future Work}
\label{sec:conclusion}
This paper presented an empirical study on residual fault prediction in Python software, combining software metrics with traditional ML and modern DL models. We constructed a balanced dataset of validated residual and non-residual faults to enable reliable evaluation. Our findings show that residual faults are primarily associated with historical and structural properties of the code. Process and coupling metrics consistently capture this signal. Supervised metric-based models achieve high recall but still produce false positives, making them practical safety filters for testing prioritization rather than definitive decision tools.
Metrics and code embeddings capture complementary information, yet straightforward hybrid models do not consistently improve performance under cross-project settings, motivating more integration strategies.

Future work will focus on extending this approach to other programming languages and evaluating its applicability to large-scale, industrial software systems with complex development histories and proprietary constraints.



\IEEEtriggeratref{63} 

\bibliographystyle{ieeetr}
\small{
    \bibliography{biblio}
}

\end{document}